\documentclass[11pt]{article}

\usepackage{amssymb}
\usepackage{graphicx}
\usepackage{youngtab}
\usepackage{chet}

\newcommand{\veps}{\varepsilon}
\newcommand{\lsp}{\hspace{1pt}}
\newcommand{\llsp}{\hspace{0.5pt}}
\newcommand{\lnsp}{\hspace{-1pt}}
\newcommand{\llnsp}{\hspace{-0.5pt}}
\newcommand{\cO}{\mathcal{O}}
\newcommand{\vV}{{\llnsp\vec{\llsp V}}}
\newcommand{\vT}{{\lnsp\vec{\llsp T}}}

\DeclareMathOperator{\Sym}{Sym}

\definecolor{darkblue}{rgb}{0.1,0.1,0.7}
\hypersetup{colorlinks,
           linkcolor={darkblue},
           citecolor={darkblue},
           urlcolor={darkblue}
}

\allowdisplaybreaks

\date{October 2018}

\preprint{CERN-TH-2018-226}

\title{Bootstrapping Mixed Correlators in\\\vspace{5pt}
Three-Dimensional Cubic Theories}

\author{Stefanos R.\ Kousvos$^a$ and Andreas Stergiou$^{b,c}$}

\affiliation{$^a$ITCP and Department of Physics, University of Crete, 700
13 Heraklion, Greece\\
$^b$Theoretical Physics Department, CERN, 1211 Geneva 23, Switzerland\\
$^c$Theoretical Division, MS B285, Los Alamos National Laboratory, Los
Alamos, NM 87545, USA}

\abstract{Three-dimensional theories with cubic symmetry are studied using
the machinery of the numerical conformal bootstrap. Crossing symmetry and
unitarity are imposed on a set of mixed correlators, and various aspects of
the parameter space are probed for consistency. An isolated allowed region
in parameter space is found under certain assumptions involving pushing
operator dimensions above marginality, indicating the existence of a
conformal field theory in this region. The obtained results have possible
applications for ferromagnetic phase transitions as well as structural
phase transitions in crystals. They are in tension with previous $\veps$
expansion results, as noticed already in earlier work.}

\begin{document}

\maketitle

\toc

\newsec{Introduction}
The cubic deformation of the Heisenberg model in three spacetime dimensions
is of paramount importance for the critical behavior of systems as simple
as magnets like Fe or Ni. In these, as well as other systems, the cubic
deformation is allowed in the context of the Landau theory of phase
transitions, and thus its effects need to be taken into account in order to
find the fixed point to which the flow is driven at low energies. In the
past, this has been addressed mainly with perturbative methods like the
$\veps$ expansion~\cite{Pelissetto:2000ek}, while Monte Carlo simulations
have been very limited~\cite{Caselle:1997gf}.  The objective in those
studies was to find out which fixed point is the stable one under
particular deformations.  In this work we study theories with cubic
symmetry using the numerical conformal bootstrap~\cite{Poland:2018epd}. The
cubic group, $C_3$, is a subgroup of the orthogonal group $O(3)$; it can be
written as a semi-direct product, $C_3=\mathbb{Z}_2{\!}^3\rtimes S_3$, or a
direct product, $C_3=S_4\times\mathbb{Z}_2$, where $S_n$ is the symmetric
group. In crystallographic notation it is the group $O_h$. Our analysis is
based just on the presence of cubic symmetry and unitarity, and does not
assume a Landau--Ginzburg description of the fixed point, as is the case
when perturbative methods are used.

One of the most widely-used strategies in the numerical conformal bootstrap
is to make an assumption about the scaling dimension of an operator (the
external operator) and obtain a bound on the scaling dimension of another
operator (the exchanged operator) that appears in the operator product
expansion (OPE) of the first operator with itself. As in other examples, in
the case of theories with cubic symmetry it is natural to first consider
the order-parameter operator $\phi_i, i=1,2,3$.  This operator has lowest
possible dimension $1/2$ consistently with unitarity, and it furnishes a
three-dimensional irreducible representation (irrep) of $C_3$. Its OPE with
itself takes the schematic form
\eqn{\phi_i\times\phi_j\sim \delta_{ij}\lsp S + X_{(ij)} + Y_{(ij)} +
A_{[ij]}\,,}[phiphiOPE]
where $S$ is the one-dimensional singlet irrep, $X$ a two-dimensional
symmetric irrep, $Y$ a three-dimensional symmetric irrep, and $A$ a
three-dimensional antisymmetric irrep. When we view $C_3$ as a subgroup of
$O(3)$, the irreps $X$ and $Y$ stem from the traceless-symmetric irrep of
$O(3)$, which is reducible under the action of $C_3$. There is a
$\mathbb{Z}_2$ symmetry under which $\phi_i$ is charged, so the operators
in the right-hand side of \phiphiOPE are all $\mathbb{Z}_2$-even.

\begin{figure}[ht]
  \centering
  \includegraphics{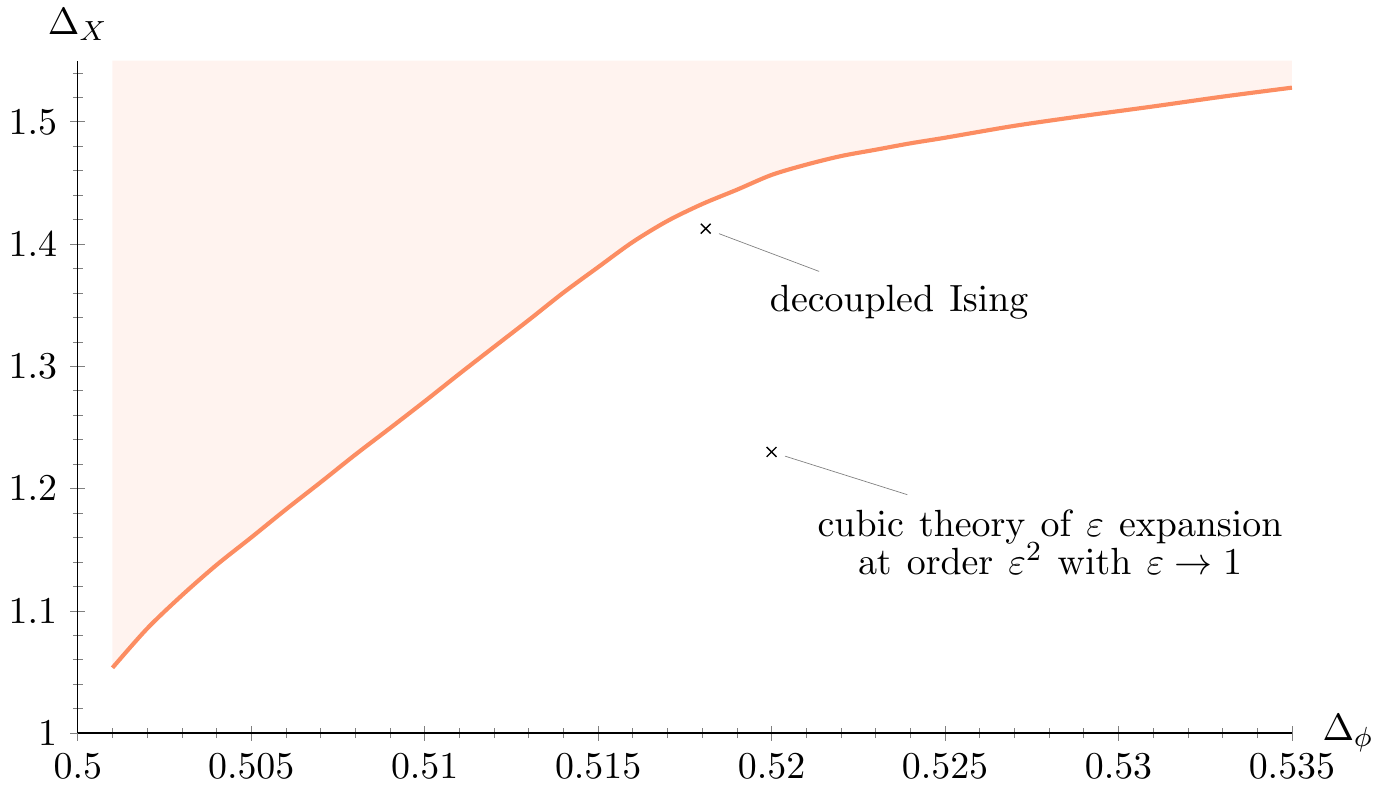}
  \caption{Upper bound on the dimension of the first $X$ operator in the
  $\phi_i\times\phi_j$ OPE. The red area is excluded.}
  \label{fig:Delta_X}
\end{figure}

In a recent paper by one of the authors~\cite{Stergiou:2018gjj}, a plot on
the dimension of the first $X$ operator in the $\phi_i\times\phi_j$ OPE was
obtained, which showed a change in slope of the boundary curve; see
Fig.~\ref{fig:Delta_X}.
Such a feature, commonly referred to as a ``kink'', has been seen in other
examples to appear due to the presence of a known conformal field theory
(CFT) at that location in parameter space, e.g.~the Ising
model~\cite{ElShowk:2012ht, El-Showk:2014dwa} and the $O(N)$
models~\cite{Kos:2013tga}. In other words, the parameters obtained when
saturating the kink have been found to be operator dimensions of a CFT. The
aim of this paper is to examine the possibility that the kink in
Fig.~\ref{fig:Delta_X} might also correspond to the location of an actual
cubic symmetric CFT, and a solution to the crossing equation that is not an
artifact of the numerics. Note that a bound on the dimension of the first
singlet scalar $S$ in the $\phi\times\phi$ OPE was also obtained
in~\cite{Stergiou:2018gjj}, but it was identical to the one obtained in the
$O(3)$ case. This limits the utility of that bound, for saturating it puts
us on the $O(3)$ solution, with the cubic one somewhere in the allowed
region. However, the coincidence of the bounds still carries some useful
information, namely that the first singlet scalar in cubic theories has
dimension lower than that in theories with $O(3)$ symmetry.

In Fig.~\ref{fig:Delta_X} we point out the position of the decoupled Ising
model, a known CFT with cubic symmetry that lies in the allowed region of
the bound.\foot{The decoupled Ising model arises simply by taking $N$
copies of the Ising model. In our case $N=3$. Each Ising model has a
$\mathbb{Z}_2$ symmetry and permuting the Ising models results in the group
$C_N=\mathbb{Z}_2{\!}^N\rtimes S_N$ for the decoupled Ising theory.} We
would like to emphasize here that this theory, in which
$\Delta_X=\Delta_\epsilon\approx1.4126$, where $\epsilon$ is the first
$\mathbb{Z}_2$-even scalar operator in the Ising model, does not saturate
the bound. Although it lies very close to it, its distance from the bound
is numerically significant; see \cite[Fig.~6]{Stergiou:2018gjj}. The bound
presented in Fig.~\ref{fig:Delta_X} has essentially converged to the
optimal one, as was verified in~\cite{Stergiou:2018gjj} by increasing the
numerical complexity of the algorithms and observing that the bound did not
get stronger.  Furthermore, an analysis of the spectrum along the bound
yielded results inconsistent with the spectrum of the decoupled Ising
model. We also include the location of the cubic theory of the $\veps$
expansion at order $\veps^2$ using results of~\cite{Dey:2016mcs}. Assuming
that higher orders and resummations do not change this location
significantly, our assumption below that $\Delta_X$ lies on the bound of
Fig.~\ref{fig:Delta_X} excludes from our subsequent plots the possibility
that our cubic theory is that of the $\veps$ expansion.

\begin{figure}[ht]
  \centering
  \includegraphics{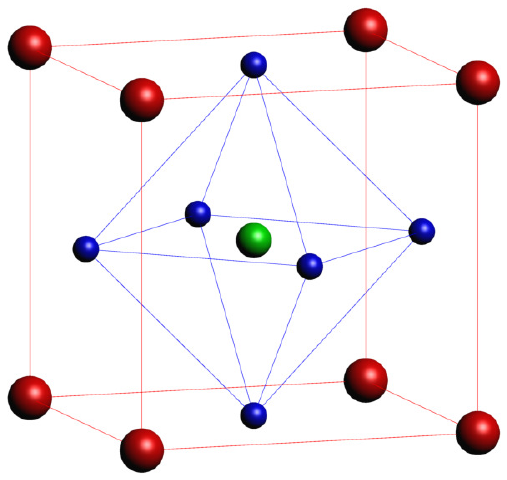}
  \caption{The perovskite structure of SrTiO$_3$. Sr is red, Ti is green,
  and O$_3$ is blue.}
  \label{fig:perovskite}
\end{figure}

\begin{figure}[ht]
  \centering
  \includegraphics{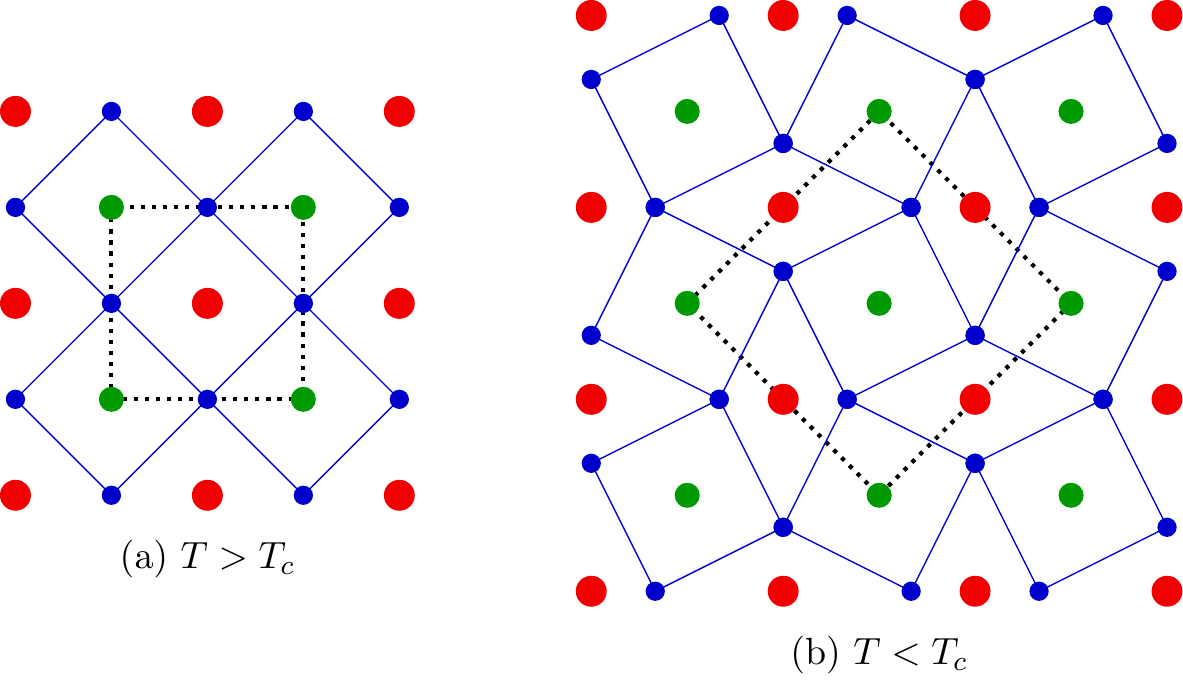}
  \caption{The crystallographic structure of SrTiO$_3$ in a top-down view
  above (a) and below (b) the critical transition temperature $T_c\approx
  100\text{ K}$. The unit cell is highlighted by the dotted line. The unit
  cell of the distorted phase is enlarged by $\sqrt{2}\times\sqrt{2}\times
  2$ relative to the undistorted phase, since two oxygen octahedra on top
  of each other rotate in opposite directions. The crystal system is cubic
  in the undistorted and tetragonal in the distorted phase.}
  \label{fig:cubic_to_tetragonal}
\end{figure}

Besides cubic magnets, CFTs with cubic symmetry have potential relevance
for structural phase transitions~\cite{Cowley, Bruce, Landau:1980mil}.
These are continuous phase transitions in which the crystallographic
structure of a crystal changes at a specific temperature, with the
high-temperature, undistorted phase having a symmetry that is broken in the
low-temperature, distorted phase. In the cubic-to-tetragonal phase
transition of SrTiO$_3$ (strontium titanate)~\cite{PhysRevLett.26.13,
RISTE19711455, PhysRevLett.28.503, PhysRevB.7.1052, CowShap}, whose
perovskite structure is seen in Fig.~\ref{fig:perovskite},
the situation is described in Fig.~\ref{fig:cubic_to_tetragonal}; the cubic
symmetry of the undistorted phase is reduced due to the transition to the
tetragonal crystallographic system below the critical temperature.
Since the cubic deformation is allowed in the undistorted phase, CFTs with
cubic symmetry and three-dimensional order parameters may be relevant for
the critical behavior of cubic systems that undergo structural phase
transitions.

With these motivations in mind we undertake in this work the study of a
system of correlation functions involving $\phi$ and $X$ using the
numerical bootstrap. More specifically, we analyze crossing and unitarity
constraints on the correlators $\langle\phi\phi\phi\phi\rangle$,
$\langle\phi\phi X\!X\rangle$ and $\langle X\!X\!X\!X\rangle$. We assume
throughout that the dimension of $X$ saturates the bound of
Fig.~\ref{fig:Delta_X}. Our considerations follow the logic described
in~\cite{Kos:2014bka, Kos:2015mba}.

When considering mixed correlators one has to analyze more OPEs besides
\phiphiOPE. In our case this consists of the OPEs $\phi\times X$ and
$X\times X$. The group theory required to understand these OPEs as well as
the decomposition of the various four-point functions under the cubic group
will be presented in detail below. Crossing symmetry leads to a system of
thirteen crossing equations which are analyzed using standard
algorithms~\cite{Behan:2016dtz, Simmons-Duffin:2015qma}.

Our numerical results show that there exists an isolated region in
parameter space, consistent with crossing and unitarity, obtained by making
assumptions of irrelevance (in the RG sense) of the second operators in the
singlet and two-dimensional irreps, called $S'$ and $X'$, respectively.
Note that these operators also appear in the OPE $\phi_i\times\phi_j$
analyzed in~\cite{Stergiou:2018gjj}, but crossing symmetry imposed only on
the correlator $\langle\phi\phi\phi\phi\rangle$ is not enough to give us
the isolated region found with the mixed correlators.  As we will see, the
essential extra constraint used in the mixed-correlator bootstrap is the
equality of certain OPE coefficients. The importance of using equality of
OPE coefficients has already been seen in the case of the $O(N)$
models~\cite{Kos:2015mba}.

Our results for the critical exponents $\beta=\Delta_\phi/(3-\Delta_S)$ and
$\nu=1/(3-\Delta_S)$ in the obtained isolated allowed region, where
$\Delta_\phi= 0.518\pm 0.001$ and $\Delta_S=1.317\pm0.012$, are
\eqn{\beta=0.308\pm0.002\,,\qquad \nu=0.594\pm0.004\,.}[crexp]
Based on \crexp we suggest that there exists a previously-unknown CFT that
is relevant for structural phase transitions, as described above, where
\cite{PhysRevLett.26.13} and \cite{PhysRevLett.28.503} give the
measurements
\eqn{\beta=0.33\pm0.02\,,\qquad \nu=0.63\pm0.07\,,}[meas]
respectively. Our suggestion is also based on the presence of cubic
symmetry and a three-dimensional order parameter. The critical exponent
$\beta$ in \meas has also been reported in the ferromagnetic phase
transition of EuS (europium sulfide)~\cite{PhysRevLett.14.71}.  Perhaps
other physical systems belong to the same universality class.  Experiments
that can shrink the critical exponents' error margins would be crucial in
testing our suggestion.

This paper is organized as follows. In the next section we describe in
detail results for the OPEs and four-point functions that are necessary for
our analysis, and list the final crossing equation we use in our numerical
explorations. In section \ref{SECres} we present our results, and we
conclude in section \ref{SECconc}.

\newsec{OPEs, four-point functions, and crossing equations}
In this section we will analyze in detail the OPEs and the four-point
functions we use in our bootstrap analysis. The consequences of cubic
symmetry for OPEs and four-point functions of operators transforming in
irreps of $C_3$ have not been explored very much in the literature, so we
will attempt to provide a self-contained treatment.

The group $C_3$ has ten irreps. Viewed as $S_4\times\mathbb{Z}_2$, these
are the five irreps of $S_4$ for each parity, namely the $\boldsymbol{1}$
(singlet), the $\bar{\boldsymbol{1}}$ (antisinglet), the $\boldsymbol{2}$
(diagonal), the $\boldsymbol{3}$ (off-diagonal), and the
$\bar{\boldsymbol{3}}$ (antisymmetric).  These irreps are nicely described
by the Young tableaux (see e.g.~\cite{Hamermesh})
\eqn{\Yvcentermath1
  \boldsymbol{1}\!:\; \yng(4)\;,\quad
  \bar{\boldsymbol{1}}\!:\; \yng(1,1,1,1)\;,\quad
  \boldsymbol{2}\!:\; \yng(2,2)\;,\quad
  \boldsymbol{3}\!:\; \yng(3,1)\;,\quad
  \bar{\boldsymbol{3}}\!:\; \yng(2,1,1)\;.\quad
}[]
The names diagonal and off-diagonal for the $\boldsymbol{2}$ and the
$\boldsymbol{3}$, respectively, stem from the location of the entries that
make them up in the traceless symmetric irrep of $O(3)$ from which they
descend. The traceless symmetric irrep of $O(3)$ is not irreducible under
the action of $C_3$, but splits into diagonal elements making up the
$\boldsymbol{2}$ and off-diagonal elements making up the $\boldsymbol{3}$.
The antisinglet irrep $\bar{\boldsymbol{1}}$ of $C_3$ is an independent
antisymmetric one-dimensional irrep.

\subsec{OPEs}
The order-parameter operator $\phi_i, i=1,2,3$ belongs to the off-diagonal
irrep and is $\mathbb{Z}_2$-odd. Its OPE with itself takes the form
\eqn{\phi_i\times\phi_j\sim\delta_{ij}S^++X_{ij}^++Y_{ij}^++A_{ij}^-\,,
}[phiphiOPEII]
where $S$ is in the singlet, $X_{ij}$ in the diagonal, $Y_{ij}$ in the
off-diagonal, and $A_{ij}$ in the antisymmetric irrep of $C_3$.  $X_{ij}$
and $Y_{ij}$ are symmetric. The signs in the superscripts indicate the spin
with which these operators appear in the OPE: even ($+$) or odd ($-$).

Let us note here that the off-diagonal irrep can be furnished by an operator
with one or with two indices. More specifically, one can write
\eqn{Y_{ij}=\gamma_{ijk}Z_k\,,}[twotoone]
where $\gamma_{ijk}$ is a symmetric tensor with
\eqn{\gamma_{123}=\tfrac{1}{\sqrt{2}}}[]
and all other independent components equal to zero. It exists only for the
group $C_3$\foot{Strictly speaking, it exists for $S_4$, not
$S_4\times\mathbb{Z}_2=C_3$.} and not for the hypercubic groups $C_{N>3}$.
Note that both $Y$ and $Z$ in \twotoone have the same $\mathbb{Z}_2$
parity.  One can verify that
\eqn{\gamma_{ijm}\gamma_{klm}=-\lsp\delta_{ijkl}
+\tfrac12(\delta_{ik}\delta_{jl}+\delta_{il}\delta_{jk})\,,}[gamgam]
where $\delta_{ijkl}$ is one if $i=j=k=l$ and zero otherwise.

They are perhaps unfamiliar, so it may be useful to give here the global
symmetry structure of the two-point functions of operators in the diagonal
and off-diagonal irreps of $C_3$. For completeness, we include the global
symmetry structure of the two-point function of antisymmetric irreps:
\threeseqn{\langle X_{ij}X_{kl}\rangle&\sim\delta_{ijkl
}-\tfrac13\delta_{ij}\delta_{kl}\,,}[twoptX]
{\langle Y_{ij}Y_{kl}\rangle&\sim-\lsp\delta_{ijkl}
+\tfrac12(\delta_{ik}\delta_{jl}+\delta_{il}\delta_{jk})\,,}[twoptY]
{\langle A_{ij}A_{kl}\rangle&\sim-\tfrac12(\delta_{ik}\delta_{jl}
-\delta_{il}\delta_{jk})\,.}[twoptA][twopt]
As we see $\langle Y_{ij}Y_{kl}\rangle=\gamma_{ijm}\gamma_{kln}\langle Z_m
Z_n\rangle\sim\gamma_{ijm}\gamma_{kln}\delta_{mn}$, which then correctly
reproduces \twoptY due to \gamgam.

Another OPE we need for our analysis is that of $\phi_i$ with $X_{jk}$,
which takes the form
\eqn{\phi_i\times X_{jk}\sim(\delta_{ijkl}
-\tfrac13\delta_{il}\delta_{jk})\lsp Y^{\prime\lsp\pm}_l
+\delta_{jkl[m}\gamma_{n]li}\lsp A^{\prime\lsp\pm}_{mn}\,.}[phiXOPE]
The operators in the right-hand side of \phiXOPE are $\mathbb{Z}_2$-odd and
$Y'$ transforms in the off-diagonal irrep.\foot{We hope the notation with
the primes indicating these $\mathbb{Z}_2$-odd irreps will not cause
confusion with the earlier notation of primes indicating the second
operators in particular irreps.} Finally, we need the OPE of $X_{ij}$ with
itself. This can be written in the form
\eqn{X_{ij}\times X_{kl}\sim(\delta_{ijkl}
-\tfrac13\delta_{ij}\delta_{kl})S^++\zeta_{ijkl}\lsp\bar{S}^-
+(\delta_{ijklmn}-\tfrac13(\delta_{ij}\delta_{klmn} +
\delta_{kl}\delta_{ijmn})+\tfrac{1}{9}\delta_{ij}\delta_{kl}\delta_{mn})
X^+_{mn}\,,}[XXOPE]
where $\delta_{ijklmn}$ is one if $i=j=k=l=m=n$ and zero otherwise, and
\eqn{\zeta_{ijkl}=\delta_{i1}\delta_{j1}(\delta_{k2}\delta_{l2}
-\delta_{k3}\delta_{l3})-\delta_{i2}\delta_{j2}(\delta_{k1}\delta_{l1}
-\delta_{k3}\delta_{l3}) + \delta_{i3}\delta_{j3}(\delta_{k1}\delta_{l1}
-\delta_{k2}\delta_{l2})\,,}[]
which is traceless in $i,j$ and $k,l$ and antisymmetric under
$ij\leftrightarrow kl$. The operators in the right-hand side of \XXOPE are
$\mathbb{Z}_2$-even. The tensor in the last term in \XXOPE comes from
$(\delta_{ijmn}-\frac13\delta_{ij}\delta_{mn})
(\delta_{klmp}-\frac13\delta_{kl}\delta_{mp})X_{np}^+$. Note that, although
not necessary, we keep the $\frac19\delta_{ij}\delta_{kl}\delta_{mn}$
contribution despite the fact that $\delta_{mn}X_{mn}^+=0$. This way the
tensor is traceless in $i,j$ and $k,l$ even before we use the tracelessness
of $X_{mn}^+$.

\subsec{Four-point functions}
With the OPEs \phiphiOPEII, \phiXOPE and \XXOPE in hand we can now proceed
to the analysis of the four-point functions relevant for our bootstrap
analysis. The strategy we employ is to expand the four-point functions in a
basis of linearly independent invariant projectors. Since we know all
required OPEs, this is a simple exercise, although sufficient care is
required in order to identify relations among particular combinations of
tensors so that we end up with linearly-independent crossing equations. We
will present our results in the $12\rightarrow34$ channel.

The four-point function $\langle\phi_i\phi_j\phi_k\phi_l\rangle$ has
already been analyzed in detail in~\cite{Stergiou:2018gjj, Rong:2017cow}.
We have
\eqna{x_{12}^{2\Delta_\phi}x_{34}^{2\Delta_\phi}
\langle\phi_i(x_1)\phi_j(x_2)\phi_k(x_3)\phi_l(x_4)\rangle&=
\sum_{S^+}\,\lambda_{\phi\phi\cO_{\llnsp S}}^2
P^{\phi\phi;\phi\phi}_{1\; ijkl}
\lsp g_{\Delta,\lsp\ell}^{\phi\phi;\phi\phi}(u,v)
+\sum_{X^+}\lambda_{\phi\phi\cO_{\!X}}^2
P^{\phi\phi;\phi\phi}_{2\; ijkl}\lsp
g_{\Delta,\lsp\ell}^{\phi\phi;\phi\phi}(u,v)\\
&\!+\sum_{Y^+}\lambda_{\phi\phi\cO_Y}^2
P^{\phi\phi;\phi\phi}_{3\; ijkl}\lsp
g_{\Delta,\lsp\ell}^{\phi\phi;\phi\phi}(u,v)
-\sum_{A^{\lnsp-}}\lambda_{\phi\phi\cO_{\!A}}^2
P^{\phi\phi;\phi\phi}_{4\; ijkl}\lsp
g_{\Delta,\lsp\ell}^{\phi\phi;\phi\phi}(u,v)\,,
}[phiphiphiphi]
where we use the conventions of~\cite{Behan:2016dtz} for the conformal
block, and the projectors are given by
\eqn{\begin{gathered}
P_{1\;ijkl}^{\phi\phi;\phi\phi}=\delta_{ij}\delta_{kl}\,,\qquad
P_{2\;ijkl}^{\phi\phi;\phi\phi}=\delta_{ijkl}
-\tfrac13\delta_{ij}\delta_{kl}\,,\\
P_{3\;ijkl}^{\phi\phi;\phi\phi}=-\lsp\delta_{ijkl}
+\tfrac12(\delta_{ik}\delta_{jl}+\delta_{il}\delta_{jk})\,,\qquad
P_{4\;ijkl}^{\phi\phi;\phi\phi}=-(\delta_{ik}\delta_{jl}
-\delta_{il}\delta_{jk})\,.\\
\end{gathered}}[phiphiphiphiProj]

Note that, strictly speaking, projectors should satisfy
\eqn{P_{I\,ijmn}P_{J\,nmkl}=P_{I\,ijkl}\lsp\delta_{IJ}\,,\qquad
\sum_{I}P_{I\,ijkl}=\delta_{il}\delta_{jk}\,,\qquad
P_{I\,ijkl}\lsp\delta_{il}\delta_{jk}=d_r^{(I)}\,,}[projeq]
where $d_r^{(I)}$ is the dimension of the representation indexed by $I$.
However, the tensors in \phiphiphiphiProj have been rescaled by positive
factors that have been absorbed into the corresponding OPE coefficients in
\phiphiphiphi, so \projeq are not satisfied by \phiphiphiphiProj without
restoring the appropriate normalizations.

For the four-point function $\langle\phi_i\phi_jX_{kl}X_{mn}\rangle$ we
find
\eqna{x_{12}^{2\Delta_\phi}x_{34}^{2\Delta_{\!X}}
\langle\phi_i(x_1)\phi_j(x_2)X_{kl}(x_3)X_{mn}(x_4)\rangle&=
\sum_{S^+}\,\lambda_{\phi\phi\cO_{\llnsp S}}\lambda_{X\!X\cO_{\llnsp S}}
P^{\phi\phi;X\!X}_{1\; ijklmn}
\lsp g_{\Delta,\lsp\ell}^{\phi\phi;X\!X}(u,v)\\
&\quad+\sum_{X^+}\lambda_{\phi\phi\cO_{\!X}}\lambda_{X\!X\cO_{\!X}}
P^{\phi\phi;X\!X}_{2\; ijklmn}\lsp
g_{\Delta,\lsp\ell}^{\phi\phi;X\!X}(u,v)\,,}[]
where
\eqna{P^{\phi\phi;X\!X}_{1\; ijklmn}&=\delta_{ij}(\delta_{klmn}
-\tfrac13\delta_{kl}\delta_{mn})\,,\\
P^{\phi\phi;X\!X}_{2\; ijklmn}&=\delta_{ijklmn}-\tfrac13(\delta_{ij}
\delta_{klmn}+\delta_{kl}\delta_{ijmn}+\delta_{mn}\delta_{ijkl})
+\tfrac29\delta_{ij}\delta_{kl}\delta_{mn}\,.}[]

The four-point function $\langle\phi_iX_{jk}\phi_lX_{mn}\rangle$ takes the
form
\eqna{(x_{12}x_{34})^{\Delta_\phi+\Delta_{\!X}}
\Big(\frac{x_{13}}{x_{24}}\Big)^{\Delta_\phi-\Delta_{\!X}}
\langle\phi_i(x_1)X_{jk}(x_2)\phi_l(x_3)X_{mn}(x_4)\rangle&=
\sum_{Y^{\prime\lsp\pm}}\,\lambda_{\phi X\cO_{Y'}}^2
P^{\phi X\llnsp;\phi X}_{1\; ijklmn}
\lsp g_{\Delta,\lsp\ell}^{\phi X\llnsp;\phi X}(u,v)\\
&\quad+\sum_{A^{\prime\lsp\pm}}\lambda_{\phi X\cO_{\!A'}}^2
P^{\phi X\llnsp;\phi X}_{2\; ijklmn}\lsp
g_{\Delta,\lsp\ell}^{\phi X\llnsp;\phi X}(u,v)\,,
}[]
where
\eqna{P^{\phi X\llnsp;\phi X}_{1\; ijklmn}&=\delta_{ijklmn}
-\tfrac13(\delta_{jk}\delta_{ilmn}+\delta_{mn}\delta_{ijkl})
+\tfrac19\delta_{il}\delta_{jk}\delta_{mn}\,,\\
P^{\phi X\llnsp;\phi X}_{2\; ijklmn}&=-\lsp\delta_{ijklmn}+\tfrac13(2\lsp
\delta_{il}\delta_{jkmn}+\delta_{jk}\delta_{ilmn}+\delta_{mn}\delta_{ijkl})
-\tfrac13\delta_{il}\delta_{jk}\delta_{mn}\,.}[]
Note, here, the relation
\eqn{\gamma_{ijk}\gamma_{lmn}=6\lsp\delta_{ijklmn}-3\Sym_{ijk}\Sym_{lmn}(3\lsp\delta_{il}\delta_{jkmn}
-\delta_{il}\delta_{jm}\delta_{kn})\,,}[]
where $\Sym_{i_1\ldots i_n}$ symmetrizes in $i_1,\ldots,i_n$ and divides by
$n!$.

We also note here the result for the four-point function
$\langle\phi_iX_{jk}X_{lm}\phi_n\rangle$, which takes the form
\eqna{(x_{12}x_{34})^{\Delta_\phi+\Delta_{\!X}}
\Big(\frac{x_{14}^2}{x_{13}x_{24}}\Big)^{\Delta_\phi-\Delta_{\!X}}
\langle\phi_i(x_1)X_{jk}(x_2)X_{lm}(x_3)\phi_{n}(x_4)\rangle&=\\
&\hspace{-16pt}
\sum_{Y^{\prime\lsp\pm}}\,\lambda_{\phi X\cO_{Y'}}^2\lsp (-1)^\ell
P^{\phi X\llnsp;X\llnsp\phi}_{1\; ijklmn}
\lsp g_{\Delta,\lsp\ell}^{\phi X\llnsp;X\llnsp\phi}(u,v)\\
&\hspace{-22pt}+\sum_{A^{\prime\lsp\pm}}
\lambda_{\phi X\cO_{\!A'}}^2\lsp(-1)^\ell
P^{\phi X\llnsp;X\llnsp\phi}_{2\; ijklmn}\lsp
g_{\Delta,\lsp\ell}^{\phi X\llnsp;X\llnsp\phi}(u,v)\,,
}[]
where
\eqn{P^{\phi X\llnsp;X\llnsp\phi}_{1\; ijklmn}=P^{\phi X\llnsp;\phi X}_{1\;
ijknlm}\,,\qquad
P^{\phi X\llnsp;X\llnsp\phi}_{2\; ijklmn}=P^{\phi X\llnsp;\phi X}_{2\;
ijknlm}\,.}[]

Finally, for the four-point function $\langle
X_{ij}X_{kl}X_{mn}X_{pq}\rangle$ we may write
\eqna{x_{12}^{2\Delta_{\!X}}x_{34}^{2\Delta_{\!X}}\langle
X_{ij}(x_1)X_{kl}(x_2)X_{mn}(x_3)X_{pq}(x_4)\rangle&=
\sum_{S^+}\,\lambda_{X\!X\cO_{\llnsp S}}^2
P^{X\!X\llnsp;X\!X}_{1\; ijklmnpq}
\lsp g_{\Delta,\lsp\ell}^{X\!X\llnsp;X\!X}(u,v)\\
&\quad-\sum_{\bar{S}^-}\,\lambda_{X\!X\cO_{\llnsp\bar{S}}}^2
P^{X\!X\llnsp;X\!X}_{2\; ijklmnpq}
\lsp g_{\Delta,\lsp\ell}^{X\!X\llnsp;X\!X}(u,v)\\
&\quad+\sum_{X^+}\lambda_{X\!X\cO_{\!X}}^2
P^{X\!X\llnsp;X\!X}_{3\; ijklmn}\lsp
g_{\Delta,\lsp\ell}^{X\!X\llnsp;X\!X}(u,v)\,,
}[]
where
\eqna{P^{X\!X\llnsp;X\!X}_{1\; ijklmnpq}&=(\delta_{ijkl}-\tfrac13
\delta_{ij}\delta_{kl})(\delta_{mnpq}-\tfrac13\delta_{mn}\delta_{pq})\,,\\
P^{X\!X\llnsp;X\!X}_{2\; ijklmnpq}&=-\tfrac13\lsp\zeta_{ijkl}\zeta_{mnpq}=
-\lsp\delta_{ijmn}\delta_{klpq}+\tfrac13(\delta_{ij}\delta_{mn}\delta_{klpq}
+\delta_{kl}\delta_{pq}\delta_{ijmn})-(mn\leftrightarrow pq)\,,\\
P^{X\!X\llnsp;X\!X}_{3\; ijklmnpq}&=-\lsp\delta_{ijkl}\delta_{mnpq}
+\delta_{ijmn}\delta_{klpq}+\delta_{ijpq}\delta_{klmn}
+\tfrac13(\delta_{ij}\delta_{kl}\delta_{mnpq}+
\delta_{mn}\delta_{pq}\delta_{ijkl})\\
&\quad-\tfrac13(\delta_{ij}\delta_{mn}\delta_{klpq}
+\delta_{kl}\delta_{pq}\delta_{ijmn})-\tfrac13(\delta_{ij}
\delta_{pq}\delta_{klmn}+\delta_{kl}\delta_{mn}\delta_{ijpq})+\tfrac19\lsp
\delta_{ij}\delta_{kl}\delta_{mn}\delta_{pq}\,.}[]
Note that the tensor $\delta_{ijklmnpq}$, which is one if
$i=j=k=l=m=n=p=q$ and zero otherwise, is not independent:
\eqna{\delta_{ijklmnpq}&=\tfrac13(\delta_{ij}\delta_{klmnpq}+
\delta_{kl}\delta_{ijmnpq}+\delta_{mn}\delta_{ijklpq}
+\delta_{pq}\delta_{ijklmn})\\
&\quad+\tfrac16(\delta_{ijkl}\delta_{mnpq}
+\delta_{ijmn}\delta_{klpq}+\delta_{ijpq}\delta_{klmn})\\
&\quad-\tfrac16(\delta_{ij}\delta_{kl}\delta_{mnpq}
+\delta_{mn}\delta_{pq}\delta_{ijkl})
-\tfrac16(\delta_{ij}\delta_{mn}\delta_{klpq}
+\delta_{kl}\delta_{pq}\delta_{ijmn})\\
&\quad-\tfrac16(\delta_{ij}\delta_{pq}\delta_{klmn}
+\delta_{kl}\delta_{pq}\delta_{ijpq})
+\tfrac16\lsp\delta_{ij}\delta_{kl}\delta_{mn}\delta_{pq}\,.}[deltaEight]
We emphasize that equations like \deltaEight are only valid for $N=3$.

\subsec{Crossing equations}
We can now impose crossing symmetry on the four-point functions involving
$\phi$ and $X$ analyzed in the previous subsection. Recall that there the
four-point functions were decomposed in the $12\rightarrow34$ channel, so
crossing symmetry requires equating those results with the decomposition of
the same four-point functions in the $14\rightarrow32$ channel. When the
dust settles we find thirteen linearly-independent crossing equations.
They can be brought to the form
\eqna{&{\sum_{S^+}}\begin{pmatrix} \lambda_{\phi\phi\cO_{\llnsp S}} &
\lambda_{X\!X\cO_{\llnsp S}} \end{pmatrix} \vT_{\!S,\Delta,\ell}
\begin{pmatrix} \lambda_{\phi\phi\cO_{\llnsp S}} \\
\lambda_{X\!X\cO_{\llnsp S}} \end{pmatrix}
+{\sum_{X^+}}\begin{pmatrix} \lambda_{\phi\phi\cO_{\!X}} &
\lambda_{X\!X\cO_{\!X}} \end{pmatrix} \vT_{\!X,\Delta,\ell}
\begin{pmatrix} \lambda_{\phi\phi\cO_{\!X}} \\
\lambda_{X\!X\cO_{\!X}} \end{pmatrix}\\
&\hspace{-0.1cm}
+{\sum_{Y^+}}\lambda_{\phi\phi\cO_{Y}}^2\!\vV_{\!Y, \Delta,\ell}
+{\sum_{A^-}}\lambda_{\phi\phi\cO_{\!A}}^2\!\vV_{\!\!A, \Delta,\ell}
+{\sum_{Y^{\prime\lsp\pm}}}\lambda_{\phi X\cO_{Y'}}^2\!
\vV_{\!Y'\!,\Delta,\ell}
+{\sum_{A^{\prime\lsp\pm}}}\lambda_{\phi X\cO_{\!A'}}^2\!
\vV_{\!\!A'\!,\Delta,\ell}
+{\sum_{{\bar{S}}^-}}\lambda_{X\!X\cO_{\!\bar{S}}}^2\!
\vV_{\!\!\bar{S},\Delta,\ell}
=0\,,
}[]
where $\vV_{\!Y, \Delta,\ell}$, $\vV_{\!\!A,\Delta,\ell}$,
$\vV_{\!Y'\!,\Delta,\ell}$, $\vV_{\!\!A'\!,\Delta,\ell}$, and
$\vV_{\!\bar{S},\Delta,\ell}$ are 13-vectors of scalar quantities, while
$\vT_{\!S,\Delta,\ell}$ and $\vT_{\!\!X,\Delta,\ell}$ are 13-vectors of
$2\times 2$ matrices. Their components are given by
\eqn{\begin{gathered}
T_{S,\Delta,\ell}^{1}=\begin{pmatrix} 0 & 0\\ 0 &
0\end{pmatrix},\quad
T_{S,\Delta,\ell}^{2}=\begin{pmatrix}
  F_{-,\Delta,\ell}^{\phi\phi;\phi\phi} & 0\\ 0 &
0\end{pmatrix},\quad
T_{S,\Delta,\ell}^{3}=\begin{pmatrix}
  F_{+,\Delta,\ell}^{\phi\phi;\phi\phi} & 0\\ 0 &
0\end{pmatrix},\quad
T_{S,\Delta,\ell}^{4}=\begin{pmatrix}
F_{-,\Delta,\ell}^{\phi\phi;\phi\phi} & 0\\ 0 &
0\end{pmatrix},\\
T_{S,\Delta,\ell}^{5}=\begin{pmatrix} 0 & 0\\ 0 &
0 \end{pmatrix},\quad
T_{S,\Delta,\ell}^{6}=\begin{pmatrix}0 & 0\\ 0 &
F_{-,\Delta,\ell}^{X\!X\llnsp;X\!X}\end{pmatrix},\quad
T_{S,\Delta,\ell}^{7}=\begin{pmatrix}0 & 0\\ 0 &
F_{+,\Delta,\ell}^{X\!X\llnsp;X\!X}\end{pmatrix},\\
T_{S,\Delta,\ell}^{8}=\begin{pmatrix} 0 &
  \tfrac12 F_{+,\Delta,\ell}^{\phi\phi;X\!X}\\
  \tfrac12F_{+,\Delta,\ell}^{\phi\phi;X\!X} &
0\end{pmatrix},\quad
T_{S,\Delta,\ell}^{9}=\begin{pmatrix} 0 &
\tfrac12 F_{+,\Delta,\ell}^{\phi\phi;X\!X} \\
\tfrac12 F_{+,\Delta,\ell}^{\phi\phi;X\!X} &
0\end{pmatrix},\quad
T_{S,\Delta,\ell}^{10\text{--}13}=\begin{pmatrix} 0 & 0 \\ 0 &
0\end{pmatrix},\\
\end{gathered}}[]
\eqn{\begin{gathered}
T_{\!X\llnsp,\Delta,\ell}^{1}=\begin{pmatrix} 0 & 0\\ 0 &
0\end{pmatrix},\,
T_{\!X\llnsp,\Delta,\ell}^{2}=\begin{pmatrix}
  -\tfrac13F_{-,\Delta,\ell}^{\phi\phi;\phi\phi} & 0\\ 0 &
0\end{pmatrix},\,
T_{\!X\llnsp,\Delta,\ell}^{3}=\begin{pmatrix}
  -\tfrac13F_{+,\Delta,\ell}^{\phi\phi;\phi\phi} & 0\\ 0 &
0\end{pmatrix},\,
T_{\!X\llnsp,\Delta,\ell}^{4}=\begin{pmatrix}
  \tfrac23F_{-,\Delta,\ell}^{\phi\phi;\phi\phi} & 0\\ 0 &
0\end{pmatrix},\\
T_{\!X\llnsp,\Delta,\ell}^{5}=\begin{pmatrix} 0 & 0\\ 0 &
  F_{-,\Delta,\ell}^{X\!X\llnsp;X\!X} \end{pmatrix},\quad
T_{\!X\llnsp,\Delta,\ell}^{6}=\begin{pmatrix}0 & 0\\ 0 &
0 \end{pmatrix}\,,\quad
T_{\!X\llnsp,\Delta,\ell}^{7}=\begin{pmatrix}0 & 0\\ 0 &
-2\lsp F_{+,\Delta,\ell}^{X\!X\llnsp;X\!X}\end{pmatrix},\\
T_{\!X\llnsp,\Delta,\ell}^{8}=\tfrac13T_{\!X\llnsp,\Delta,\ell}^{10}=
\begin{pmatrix} 0 &
  \tfrac16 F_{-,\Delta,\ell}^{\phi\phi;X\!X}\\
  \tfrac16F_{-,\Delta,\ell}^{\phi\phi;X\!X} &
0\end{pmatrix}\,,\quad
T_{\!X\llnsp,\Delta,\ell}^{9}=\tfrac13T_{\!X\llnsp,\Delta,\ell}^{11}=
\begin{pmatrix} 0 &
\tfrac16 F_{+,\Delta,\ell}^{\phi\phi;X\!X} \\
\tfrac16 F_{+,\Delta,\ell}^{\phi\phi;X\!X} &
0\end{pmatrix},\\
T_{\!X\llnsp,\Delta,\ell}^{12,13}=\begin{pmatrix} 0 & 0\\ 0 &
0\end{pmatrix},
\end{gathered}}[]
\eqn{\begin{gathered}
V_{Y,\Delta,\ell}^{1}=F_{-,\Delta,\ell}^{\phi\phi;\phi\phi}\,,\quad
V_{Y,\Delta,\ell}^{2}=F_{-,\Delta,\ell}^{\phi\phi;\phi\phi}
\,,\quad
V_{Y,\Delta,\ell}^{3}=-F_{+,\Delta,\ell}^{\phi\phi;\phi\phi}
\,,\quad
V_{Y,\Delta,\ell}^{4\text{--}13}=0\,,
\end{gathered}}[]
\eqn{\begin{gathered}
V_{\!A,\Delta,\ell}^{1}=F_{-,\Delta,\ell}^{\phi\phi;\phi\phi}\,,\quad
V_{\!A,\Delta,\ell}^{2}=-F_{-,\Delta,\ell}^{\phi\phi;\phi\phi}\,,\quad
V_{\!A,\Delta,\ell}^{3}=F_{+,\Delta,\ell}^{\phi\phi;\phi\phi}\,,\quad
V_{\!A,\Delta,\ell}^{4\text{--}13}=0\,,
\end{gathered}}[]
\eqn{\begin{gathered}
V_{Y'\llnsp,\Delta,\ell}^{1\text{--}7}=0\,,\quad
V_{Y'\llnsp,\Delta,\ell}^{8}=
\tfrac23V_{Y'\llnsp,\Delta,\ell}^{10}=\tfrac23(-1)^\ell
  F_{-,\Delta,\ell}^{\phi X\llnsp;X\phi}\,,\quad
V_{Y'\llnsp,\Delta,\ell}^{9}=
\tfrac23V_{Y'\llnsp,\Delta,\ell}^{11}=\tfrac23(-1)^{\ell+1}
F_{+,\Delta,\ell}^{\phi X\llnsp;X\phi}\,,\\
V_{Y'\llnsp,\Delta,\ell}^{12}=
  F_{-,\Delta,\ell}^{\phi X\llnsp;\phi X}\,,\quad
V_{Y'\llnsp,\Delta,\ell}^{13}=0\,,
\end{gathered}}[]
\eqn{\begin{gathered}
V_{A'\llnsp,\Delta,\ell}^{1\text{--}9}=0\,,\quad
V_{A'\llnsp,\Delta,\ell}^{10}=(-1)^{\ell+1}
  F_{-,\Delta,\ell}^{\phi X\llnsp;X\phi}\,,\quad
V_{A'\llnsp,\Delta,\ell}^{11}=(-1)^\ell
F_{+,\Delta,\ell}^{\phi X\llnsp;X\phi}\,,\\
V_{A'\llnsp,\Delta,\ell}^{12}=0\,,\quad
V_{A'\llnsp,\Delta,\ell}^{13}=
F_{+,\Delta,\ell}^{\phi X\llnsp;\phi X}\,,
\end{gathered}}[]
and
\eqn{\begin{gathered}
V_{\!\bar{S},\Delta,\ell}^{1\text{--}4}=0\,,\quad
V_{\!\bar{S},\Delta,\ell}^5=-V_{\!\bar{S},\Delta,\ell}^6=
  F_{-,\Delta,\ell}^{X\!X\llnsp;X\!X}\,,\quad
V_{\!\bar{S},\Delta,\ell}^7=
  F_{+,\Delta,\ell}^{X\!X\llnsp;X\!X}\,,\quad
V_{\!\bar{S},\Delta,\ell}^{8\text{--}13}=0\,.
\end{gathered}}[]
In these equations we use the standard notation
\eqn{F_{\pm,\Delta,\ell}^{\cO_1\cO_2;\cO_3\cO_4}(u,v)=
v^{\frac12(\Delta_2+\Delta_3)}
g_{\Delta,\ell}^{\cO_1\cO_2;\cO_3\cO_4}(u,v)
\pm(u\leftrightarrow v)\,,}[]
where the dependence of the conformal block
$g_{\Delta,\ell}^{\cO_1\cO_2;\cO_3\cO_4}(u,v)$ on the dimensions $\Delta_a$
of the operators $\cO_a$, $a=1,\ldots,4$, comes only through the
combinations $\Delta_1-\Delta_2$ and $\Delta_3-\Delta_4$.

\newsec{Results}[SECres]
In this section we present the results of our numerical exploration of
$\phi$-$X$ mixed correlators in unitary theories with cubic symmetry.
Parameter choices in \texttt{PyCFTBoot} are as follows: $\texttt{nmax=7}$,
$\texttt{mmax=5}$, $\texttt{kmax=32}$. We include spins up to
$\ell_{\text{max}}=26$. For \texttt{SDPB} we use the options
\texttt{--findPrimalFeasible} and \texttt{--findDualFeasible},\foot{With
these options if \texttt{SDPB} finds a primal feasible solution then the
assumed operator spectrum is allowed, while if it finds a dual feasible
solution then the assumed operator spectrum is excluded.} and we further
set $\texttt{--precision=660}$, $\texttt{--dualErrorThreshold=1e-20}$, and
default values for other parameters.

\begin{figure}[ht]
  \centering
  \includegraphics{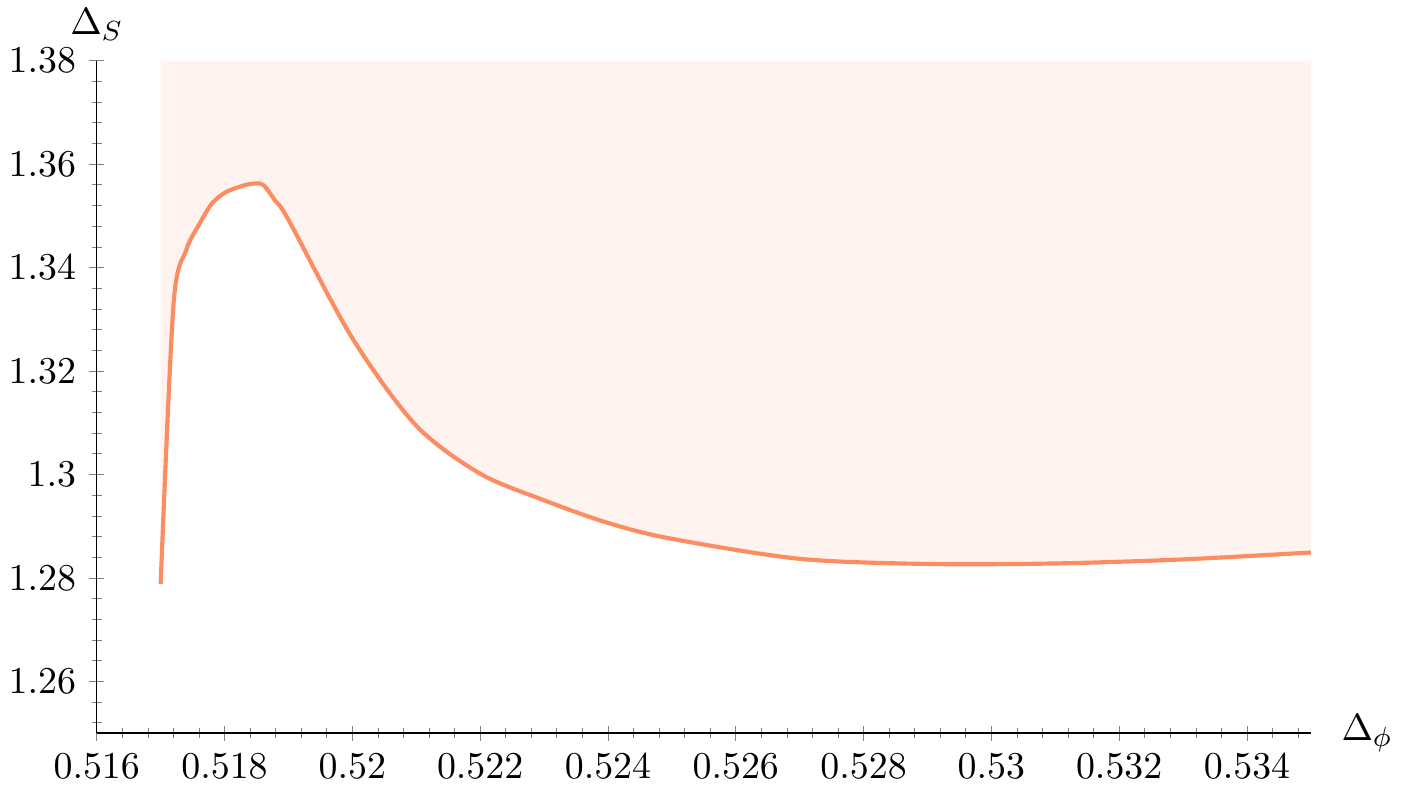}
  \caption{Upper bound on the dimension of the first singlet operator $S$.
  For this plot we assume that $\Delta_X$ lies on the bound of
  Fig.~\ref{fig:Delta_X} and we impose the gaps $\Delta_{X'}>3.0$ and
$\Delta_{\phi'}>1.0$. The red area is excluded. }
  \label{fig:Delta_S}
\end{figure}

As a first result we would like to mention that the plot of
Fig.~\ref{fig:Delta_X} remains the same even when we use the system of
$\phi$-$X$ mixed correlators. Assuming that $\Delta_X$ lies on the bound of
Fig.~\ref{fig:Delta_X}, we can obtain a bound on $\Delta_S$ using the
additional assumptions $\Delta_{X'}>3.0$ and $\Delta_{\phi'}>1.0$. We
remind the reader that $\phi$ appears in the $\phi\times X$ OPE in the
$Y^{\prime\lsp+}$ set of operators for spin zero, and the corresponding OPE
coefficient is equal to the OPE coefficient with which $X$ appears in the
$\phi\times\phi$ OPE, i.e.\ $c_{\phi\phi X}=c_{\phi X\phi}$. Unless
otherwise noted, we use this OPE coefficient equality throughout this
section. The next scalar operator in the $Y^{\prime\lsp+}$ set of operators
is here called $\phi'$, and we impose the mild gap $\Delta_{\phi'}>1.0$ in
order to ensure that the equality of the OPE coefficients we mentioned
provides an actual constraint~\cite{Kos:2015mba}.\foot{We have verified
that our results are not sensitive to the gap we choose on
$\Delta_{\phi'}$, assuming it remains between $0.6$ and $1.1$.}  The bound
is shown in Fig.~\ref{fig:Delta_S}.
The bound of Fig.~\ref{fig:Delta_S} clearly indicates a region, located
horizontally around $\Delta_\phi=0.518$, in which one could expect to find
a special solution to the crossing equations.

\begin{figure}[!ht]
  \centering
  \includegraphics{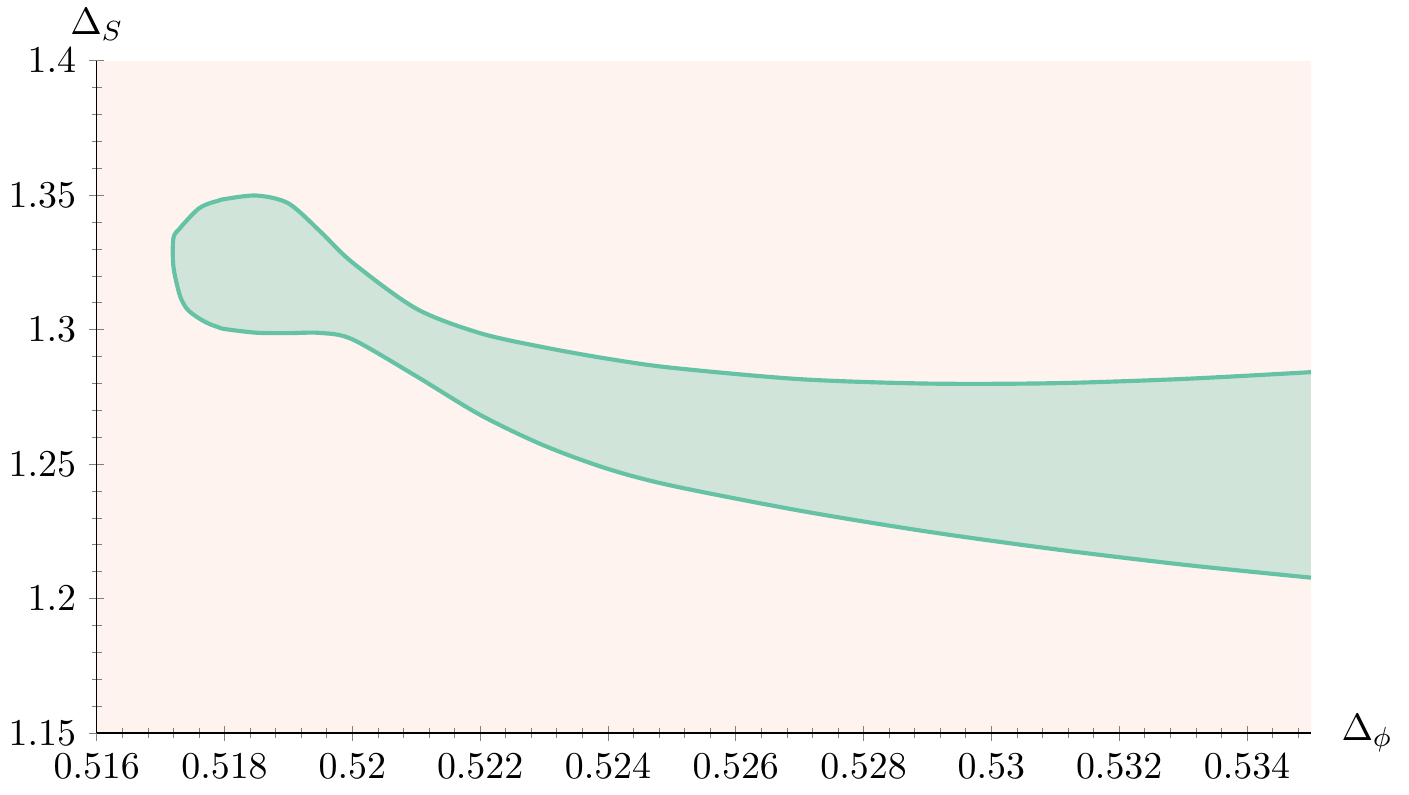}
  \caption{Allowed region, in green, for the dimension of the first singlet
  operator $S$. The red area is excluded. To obtain this plot we make the
  assumptions $\Delta_{S'},\Delta_{X'}>3.0$ and $\Delta_{\phi'}>1.0$. The
  allowed region here looks like an arm, with the narrow wrist and the
  wider fist.}
  \label{fig:Delta_S_peninsula}
\end{figure}

Still assuming that $\Delta_X$ lies on the bound of Fig.~\ref{fig:Delta_X},
we now obtain a region of allowed $\Delta_S$, with the assumptions
$\Delta_{S'}, \Delta_{X'}>3.0$, and $\Delta_{\phi'}>1.0$. Our results are
shown in Fig.~\ref{fig:Delta_S_peninsula}.
Compared to Fig.~\ref{fig:Delta_S}, the only extra assumption for the plot
of Fig.~\ref{fig:Delta_S_peninsula} is that we take $\Delta_{S'}>3.0$.
Clearly, the shape of the allowed region in
Fig.~\ref{fig:Delta_S_peninsula} suggests that something special is
happening in the region of the fist.  This is because of the fact that the
narrowing of the allowed region observed all the way to the wrist does not
continue, but instead we see a widening. This indicates that a solution to
crossing symmetry, and hence a CFT, lies in the fist of the allowed region.

\begin{figure}[ht]
  \centering
  \includegraphics{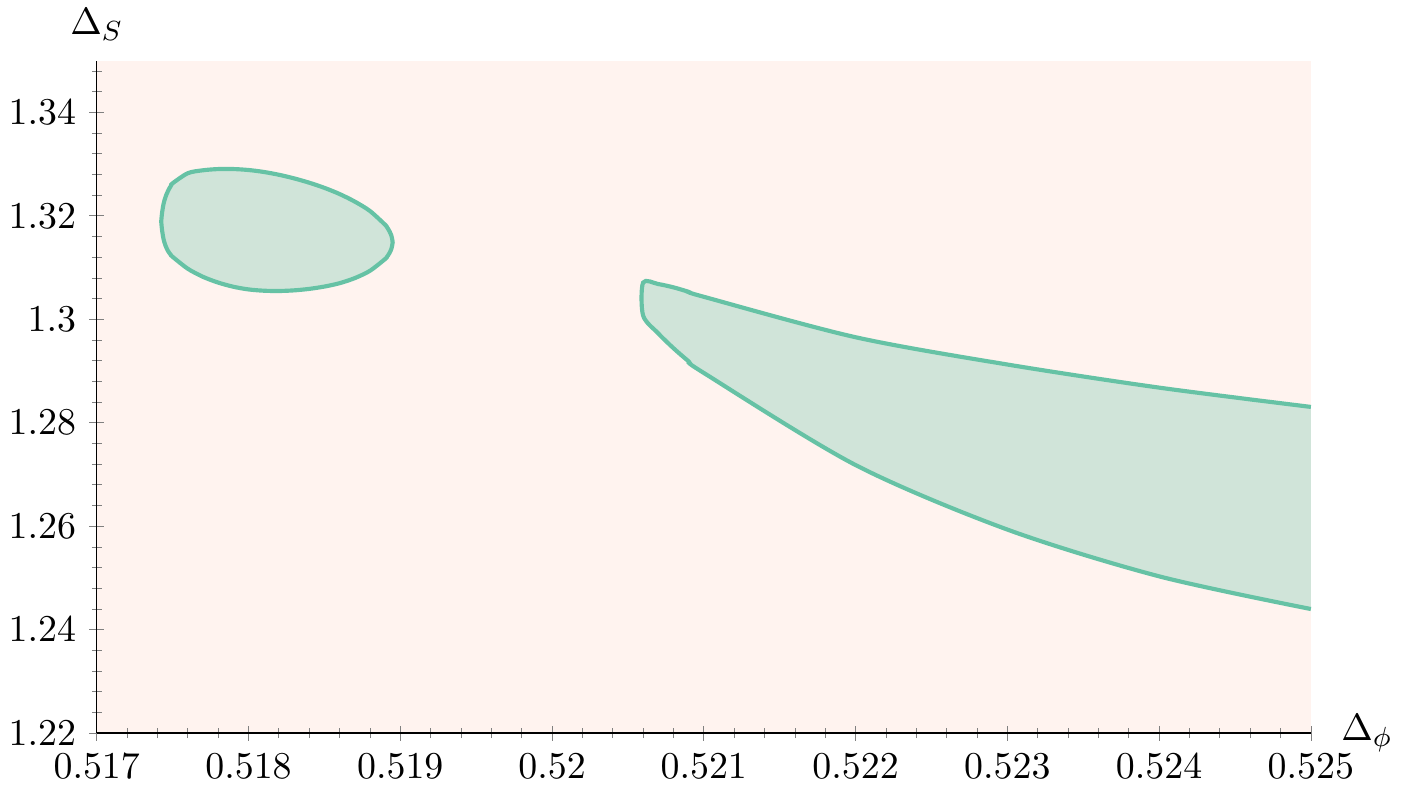}
  \caption{Allowed region, in green, for the dimension of the first singlet
  operator $S$. The red area is excluded. To obtain this plot we make the
assumptions $\Delta_{S'}>3.8$, $\Delta_{X'}>3.0$ and $\Delta_{\phi'}>1.0$.}
  \label{fig:Delta_S_island}
\end{figure}

\begin{figure}[!ht]
  \centering
  \includegraphics{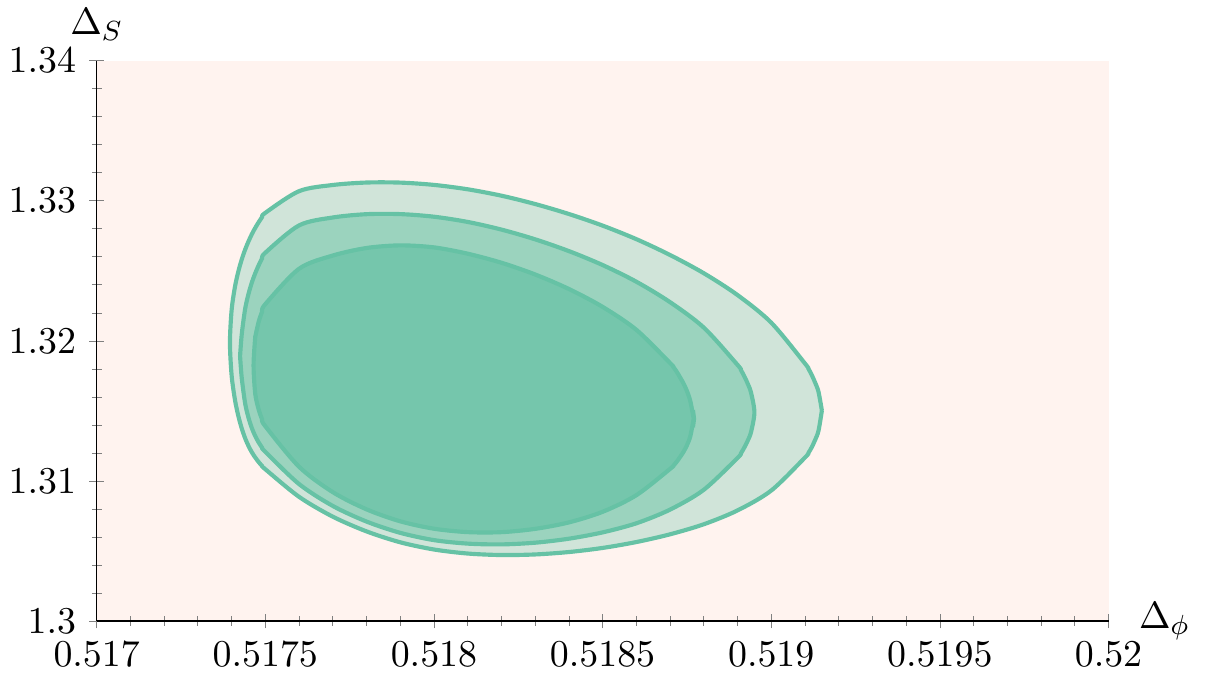}
  \caption{Allowed regions, in green, for the dimension of the first
  singlet operator $S$. The red area is excluded. Three closed allowed
  regions are plotted in green. For all of them we assume $\Delta_{X'}>3.0$
  and $\Delta_{\phi'}>1.0$. We also assume $\Delta_{S'}>3.7, 3.8, 3.9$ to
  obtain the largest, intermediate, and smallest allowed region,
  respectively.}
  \label{fig:Delta_S_islands}
\end{figure}

Let us now make the assumptions $\Delta_{S'}>3.8$, $\Delta_{X'}>3.0$, and
$\Delta_{\phi'}>1.0$. We only raised the gap of $S'$ compared to the
previous choices. We can see that, as we raise this gap, the wrist of the
allowed region in Fig.~\ref{fig:Delta_S_peninsula} narrows. With
$\Delta_{S'}>3.8$ the fist actually separates from the rest of the arm, and
we obtain an isolated allowed region! This is shown in
Fig.~\ref{fig:Delta_S_island}. Note that had we not imposed the equality
$c_{\phi\phi X}=c_{\phi X\phi}$, we would not have obtained the separation
that led to the isolated allowed region.
We would like to point out that an isolated region of roughly the same size
and shape is obtained for choices of the $\Delta_{S'}$ gap between
$\Delta_{S'}\approx 3.7$ and $\Delta_{S'}\approx 3.9$. For
$\Delta_{S'}\gtrsim 3.9$ the isolated allowed region is abruptly lost.
This is consistent with results of~\cite[Fig.~7]{Stergiou:2018gjj}. The
isolated regions for the different choices of the $\Delta_{S'}$ gap are
shown in Fig.~\ref{fig:Delta_S_islands}.

Figs.~\ref{fig:Delta_S_island} and \ref{fig:Delta_S_islands} contain the
most important results of this paper. It is natural to assert that the
isolated allowed regions seen there contain operator dimensions of an
actual strongly-coupled CFT.

\newsec{Conclusion}[SECconc]
In this work we have carried out a detailed numerical analysis of theories
with cubic symmetry in three dimensions. We analyzed a system of mixed
four-point functions, and after experimenting with assumptions on the
spectrum we managed to find an isolated region allowed by unitarity and
crossing symmetry. Based on earlier bootstrap experience, where similar
allowed regions were found around already known CFTs, we concluded that our
allowed region hosts a CFT with cubic symmetry. This CFT has only one
relevant scalar singlet operator. In experiments this would correspond to
the temperature that needs to be tuned in order to reach the critical
point.

The CFT we are studying here does not appear to be previously known. The
standard tool for looking for CFTs in $d=3$ is the $\veps$ expansion below
$d=4$. The $\veps$ expansion gives a nontrivial cubic theory in
$d=4-\veps$~\cite{Aharony:1973zz, Pelissetto:2000ek, Osborn:2017ucf}, but
as already noticed in~\cite{Stergiou:2018gjj} results for operator
dimensions obtained for that theory are in disagreement with what we find
here.\foot{The results of \cite{Stergiou:2018gjj} were obtained with a
spectrum analysis~\cite{Poland:2010wg, ElShowk:2012hu} consistent with our
results in this work.} For example, the $\veps$
expansion~\cite{Kleinert:1994td} as well as the fixed-dimension methods
used in~\cite{Carmona:1999rm} give $\Delta_{S'}$ just slightly above
marginality, which is not the case for our theory. We find the possibility
that the $\veps$ expansion produces wrong results unlikely, although we
cannot exclude it.

Another, more likely possibility, is that the theory we find here is a
theory with cubic symmetry that cannot be captured with perturbative
methods. A class of such theories can be obtained after resummations of
perturbative beta functions that lead to extra fixed points not found in
perturbation theory. As an example, these resummation methods have
suggested a non-perturbative fixed point for $O(3)\times O(2)$ frustrated
spin systems~\cite{Pelissetto:2000ne, Calabrese:2002af, Calabrese:2004nt},
evidence for which has also been found with the
bootstrap~\cite{Nakayama:2014sba}. The existence of such fixed points,
however, has been challenged in~\cite{Delamotte:2008, Delamotte:2010ba,
Delamotte:2010bb}.  Using the results of \cite{Carmona:1999rm} a
non-perturbative fixed point cannot be found for cubic theories.\foot{We
thank A.~Vichi for informing us about this fact.} It is possible that
resummations performed with higher than six-loop results would reveal
evidence for a strongly-coupled cubic CFT not seen in the $\veps$
expansion. Clearly, Monte Carlo simulations would be very helpful in
settling this issue. Note that our CFT has no relevance to cubic magnets at
the critical point, for critical exponents for those systems have been
measured and are very close to the ones predicted by the Heisenberg model.
We propose to refer to our CFT by the name ``Platonic CFT''.\foot{The
Platonic solids are well-known, as are some of their properties. In
particular, cubes are dual to octahedra and dodecahedra are dual to
icosahedra. Tetrahedra are self-dual. For $N=3$ the symmetry groups $C_3$
and $T_3\times\mathbb{Z}_2$, where $T_3=S_4$ is the symmetry group of the
tetrahedron, are isomorphic.  There is no chance of an dodecahedral CFT, at
least in the $\veps$ expansion below $d=4$, so the name ``Platonic CFT''
appears fitting.}

Future work includes enlarging the set of operators we consider in our
four-point functions. In this work we considered two operators, namely
$\phi_i$ and $X_{ij}$, but in the future we would like to analyze
numerically the system of correlators involving the scalar singlet $S$ as
well. With the full system of crossing equations we hope to be able to get
a three-dimensional isolated allowed region in the plot of allowed
dimensions of $\phi_i$, $X_{ij}$, and $S$. We also hope to be able to push
the precision in order to obtain more accurate determinations of the
critical exponents. In that regard, the so-called $\theta$-scan explored
in~\cite{Kos:2016ysd, Atanasov:2018kqw} may also lead to improvements. For
these future endeavors we would greatly benefit from faster numerical
optimization algorithms than are currently available.

%%fakesection Acknowledgments
\ack{We would like to thank C.~Behan for a modification of PyCFTBoot that
allowed us to easily impose equality of OPE coefficients. AS thanks
A.~Vichi for many illuminating discussions. We also thank H.~Osborn,
S.~Rychkov, and A.~Vichi for comments on the manuscript. SRK would like to
thank the Crete Center for Quantum Complexity and Nanotechnology for use of
the Metropolis cluster, as well as G.~Kapetanakis for computing support.
SRK would also like to thank ITCP Crete for financial support as well as
CERN-TH and AS for hospitality during a visit. The numerical computations
in this paper were run on the LXPLUS cluster at CERN and the Metropolis
cluster at the Crete Center for Quantum Complexity and Nanotechnology.}

\bibliography{bootstrapping_mixed_cubic_correlators}

\end{document}